# Micromagnetic Study of Spin Transfer driven Vortex Dipole and Vortex Quadrupole Dynamics

A. Giordano[1], V. Puliafito[1], L. Torres[2], M. Carpentieri[3], B. Azzerboni[1], and G. Finocchio[1]

[1] Department of Electronic Engineering, Industrial Chemistry and Engineering, University of Messina, c.da Di Dio, 98166 Messina, Italy
[2] Department of Applied Physics, University of Salamanca, Plaza de los Caídos, 38008 Salamanca, Spain
[3] Department of Electrical and Information Engineering, Politecnico of Bari, via Orabona 4, 70125 Bari, Italy

*Abstract*—Vortices and antivortices are typical non uniform magnetization configurations that can be achieved in spin-torque oscillators with in-plane materials. Dynamics of a vortex-antivortex pair, namely vortex dipole, were predicted and already observed. In this paper, we present a systematic micromagnetic study on that kind of dynamics in different spin-valves where the current is injected through a nano-aperture. Rotation and translation of vortex dipoles and rotation of vortex quadrupoles are shown depending on the shape and the size of the spin-valve. The origin of the different behaviors is explained within a micromagnetic framework.

## I. INTRODUCTION

NON UNIFORM magnetization configurations have been found to be the underlying cause of most of the phenomena arising in the field of magnetic nanostructures in the last years [1], [2], [3], [4], [5], [6], [7], [8]. In the spintronic applications scenario [9], [10], [11] non-uniform configurations are present not only in the micrometer sized sensors [12], or extended nano-contact spin-transfer driven oscillators (NC-STOs) [13], [14], but also in spin valves (SVs) with dimensions in the range of a hundred nanometers [15]. For studying all these devices, full micromagnetic formalism is the adequate framework for getting insight in the complex non-uniform magnetization dynamics of the nanostructure [16], [17], [18].

Injection of a direct electrical current through SVs offers the possibility to induce microwave steady-state magnetization precession by the action of the spin transfer torque [9]. In particular, rotational vortex and antivortex motion can be excited by a direct spin-polarized current as already demonstrated both experimentally and numerically in [19]. This soliton-type dynamics of the vortex-antivortex pair (vortex dipole) was also predicted by analytical models in the case of field induced dynamics [20]. Similar analytical models were proposed also for the case of spin transfer torque induced vortex dipole dynamics [21].

Here we present a more ample full micromagnetic study on the vortex dipole soliton dynamics in different kinds of sub-micrometer SVs where the current is injected through a nano-aperture of tenths of nanometers. All the features predicted by the first analytical models [20], [21] are reported, including vortex dipole rotation or translation depending on the shape (circular, elliptical, rectangular) and size of the SV and also on the field and/or current applied.

## II. MICROMAGNETIC MODEL AND COMPUTATIONAL DETAILS

We study the spin-transfer-torque induced magnetization dynamics in spin-valves due to a non localized current. The active part of the spin-valve (see Fig. 1(a)) is composed by Py($Ni_{81}Fe_{29}$)(5)/$Al_2O_3$(3.5)/Cu(8)/Py(20) (thicknesses in nm).

We have analyzed different realistic geometrical and physical parameters, but here we will point our attention on three different setups:
a) Elliptical cross section with axes of 250 nm and 150 nm and axes of 250 nm and 120 nm;
b) Circular cross section with diameter of 250 nm;
c) Rectangular cross section with dimensions 250x150 $nm^2$ and 500x300 $nm^2$.

A 40 nm diameter nanoaperture is considered in all the devices to enable the localized injection of spin-polarized currents into the SV. We will refer to the thinner Py layer as the free layer (FL) and to the thicker layer as the pinned layer (PL). We will employ a Cartesian system of coordinates in which the major axis of the ellipse is the *x* (axis *x*) and the in-plane hard axis is the *y* (axis *y*), and we will use the convention that positive current corresponds to the flow of electrons from the FL to the PL.

Our results are based on the numerical solution of the Landau-Lifshitz-Gilbert-Slonczweski (LLGS) equation:

$$\frac{d\mathbf{m}}{d\tau} = -\mathbf{m} \times \mathbf{h}_{eff} + \alpha \mathbf{m} \times \frac{d\mathbf{m}}{d\tau} + \\ - \frac{g|\mu_B|J_Z(x,y)}{e\gamma_0 M_S^2 L}\varepsilon(\vartheta)\mathbf{m}\times(\mathbf{m}\times\mathbf{m}_p) \quad (1)$$

where $\mathbf{m} = \mathbf{M}/M_s$ is the dimensionless magnetization vector ($M_s$ saturation magnetization), $\mathbf{h}_{eff} = \mathbf{H}_{eff}/M_s$ is the effective field, and $\mathbf{m}_p$ is the fixed layer magnetization. *g* is the gyromagnetic splitting factor, $\gamma_0$ the gyromagnetic ratio, $\mu_B$ the Bohr magneton and *e* the electron charge. $J_Z(x,y)$ is the current density through the nano-aperture and *L* the thickness of the dynamic layer. The damping parameter is $\alpha$ and $\varepsilon(\vartheta)$ is the spin transfer torque efficiency function. The computations are performed with our own micromagnetic code, a finite differences time domain code, details can be found in [7], [18]. For extremely long computational times or largest samples, parallelized code GPMagnet [17] has been used to optimize computational times.

Except when specifically indicated, the standard parameters used in simulations throughout all the paper were: $M_s$ = 650 kA/m, $\alpha$ = 0.01, cubic computational cells of 5 nm in side and computational time discretization of 50 fs.

The contributions included in $H_{eff}$ were: demagnetizing field, exchange field (with an exchange constant $A$ = 1.3 $10^{-11}$ J/m), Oersted field, and external field ($H_x$ = -25 mT). The initial state was free and pinned layers aligned along the same (+$x$) direction. The current is considered as uniform below the nanoaperture and zero outside.

### III. RESULTS AND DISCUSSION

In the following we will analyze the dynamics of the vortex dipole and quadrupole structures: i) vortex dipole rotation, ii) vortex dipole translation and iii) vortex quadrupole rotation.

*i) Vortex dipole rotation.*

As mentioned before, in a previous work [19] the dynamics of the vortex dipole in an elliptical device (250x150 nm$^2$) both experimentally and by means of micromagnetic simulations was found. It is useful to remind that such a vortex dipole consists of a vortex and an antivortex pair with opposite polarities which rotates counterclockwise (CCW) around the centre. Here, we will get a deeper insight in some aspects of the vortex rotation in elliptical devices showing how to control the rotation frequency by the applied current and by geometrical aspects of the device (thickness and aspect ratio). For the same elliptical section reported in [19], therefore, we present in Fig. 1(b) the current dependence of the vortex dipole rotation frequency for three different thicknesses of the FL. As it is possible to notice, the frequency decreases with increasing current. In our computations, moreover, the increasing of current leads to an increase of the distance between the vortex and antivortex cores, $d_{VA}$ (not shown). Our results, therefore, are in qualitative agreement with analytical models [20], [21] where the frequency is inversely proportional to $d_{VA}$. Quantitative comparison with analytical models is not possible since these models do not take specifically into account the magnetostatic field, although some qualitative comments were mentioned in [21]. The effect of magnetostatic field can be analyzed just by using full micromagnetics as shown in our Fig. 1(b), where the frequency increases with the thickness. For larger thicknesses, in fact, the demagnetizing energy is lower, allowing a stronger attraction of vortex and antivortex cores. This leads to a more contracted vortex dipole, i.e. lower $d_{VA}$ and higher frequency rotation.

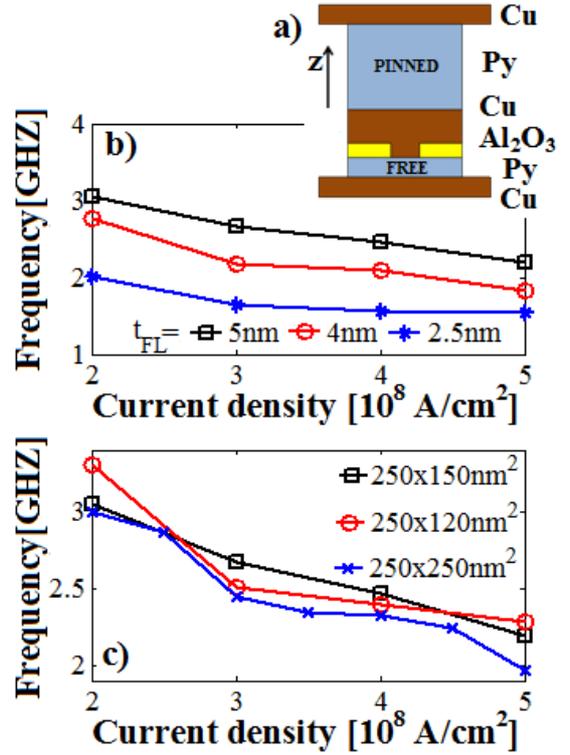

Fig. 1. (a) Schematics of the device geometry. Its cross section gets different shapes as described in the text. (b) Dependence of the rotation frequency on the current and thickness in the case of elliptical device (250x150 nm$^2$). (c) Comparison between different devices with H=25mT.

The magnetostatic field is also varied when modifying the aspect-ratio of the elliptical device or when going to a circular device. In Fig.1(c) we study the frequency-current dependence for two elliptical sections, 250x150 nm$^2$ and 250x120 nm$^2$, with a thickness of 5nm, and for the circular device (250 nm diameter). Actually, there are no evident differences in the different cases. Nevertheless, numerical simulations for the circular device show jumps of the sense of rotation of the vortex dipole. CCW rotation corresponds to a vortex with positive polarity and an antivortex with negative polarity (Fig. 2(a)), viceversa for the clockwise (CW) rotation (Fig. 2(b)). These jumps disappear by slightly increasing the current over the range illustrated in Fig. 1(c). The reason of such a behavior, that was not found out in the elliptical section devices, has to be found in the symmetry of the circular device. In this case, in fact, there is a sole field contribution that breaks the symmetry, namely the Oersted field. It introduces an energetic gap between the two states ($\Delta e$ in the schematics of Fig. 2(c)). Nonetheless, this gap is usually small and each state can jump to the other, for instance by means of the reflection of spin waves from the device borders. The increase of the current enlarges the energetic gap so leading to a stable state, CCW or CW depending on the transient. A further confirmation of this aspect was obtained by removing the Oersted field contribution. In this case, in fact, it is not possible to observe any ordered rotation of the vortex dipole, since the energetic gap is null and none of the two senses can predominate even for a short time. In the case of elliptical section device, differently, the energetic gap turns out higher due to the shape anisotropy so that only a sense of rotation is observed, depending again on the transient.

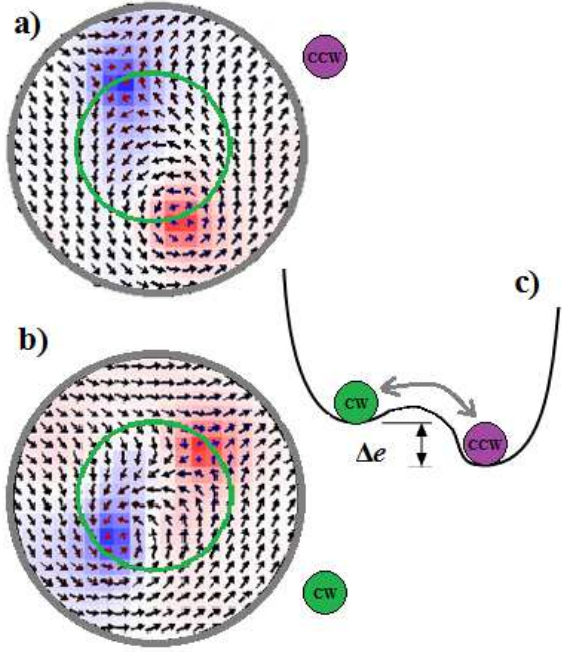

Fig. 2. Magnetization configuration in the circular device (250x250 nm$^2$), with (a) counterclockwise and (b) clockwise rotation of the vortex dipole. (c) Schematics for the different energetic levels of the two senses of rotation. $\Delta e$ depends on the Oersted field only in the case of circular device.

*ii) Vortex dipole translation*

Vortex dipole translation in the direction perpendicular to the line connecting the vortex antivortex cores was predicted in the first topological studies of vortex dipoles [22]. We have also found this behavior in our micromagnetic study in the case of rectangular devices with dimensions 250x150 nm$^2$. The vortex dipole is characterized by the same polarity of vortex and antivortex and it translates in a complicated way (see Fig. 3), changing its translation direction due to the deformation of the dipole because of the magnetostatic interaction. It has to be considered that the topological models [22] are constructed for infinite medium where the translation of the rigid dipole along its axis perpendicular direction is to expect.

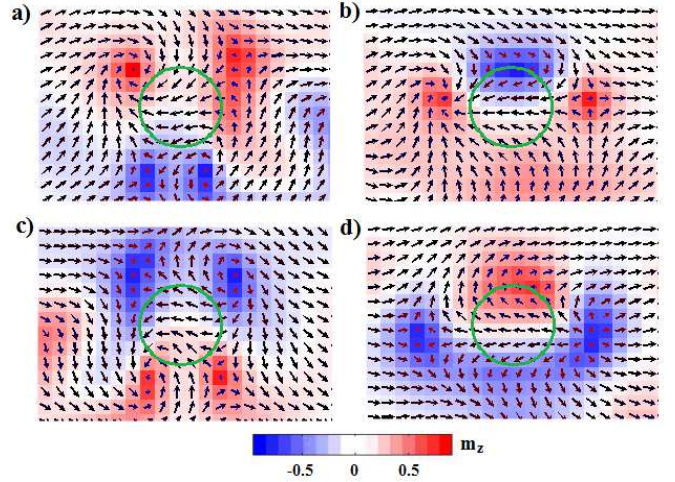

Fig. 3. Configuration of the magnetization in different time instants in the case of vortex dipole translation in a rectangular section device of 250x150 nm$^2$ (a zoomed-in area of the device is shown). Circular solid lines represent the contact area.

*iii) Vortex quadrupole rotation*

A completely different new topological structure is found when increasing the rectangular area but with the same aspect ratio, that is for dimensions 500x300 nm$^2$. As shown in Fig. 4, a vortex quadrupole stable rotation is found. Vortex quadrupoles were previously reported as intermediate states in large samples dynamics [16] but always vortex-antivortex pairs annihilated for leading to a final vortex state or vortex dipole state. Here, we report a stable oscillation of a vortex quadrupole, in the stationary rotational state, where the two vortices have the same polarity (+1) and the two antivortices have also the same polarity (-1), rotating CCW. Prior, to reach this stationary rotational state, vortex and antivortex pairs are created and expelled from the rotation area till the symmetric quadrupole is formed and stabilized.

The frequency of this quadrupole rotation can be tuned by either current and external field applied as shown in Fig. 4(a). Since no topological model is at hand for this structure, we can just make some conjectures based on the vortex dipole models. The decreasing of the frequency with current could be explained like in the dipole, an increase of current leads to an increase of the quadrupole size and the corresponding decrease in frequency. The increasing of the external field implies that most of the magnetization tends to align in the direction of that field, leading to a decrease of the quadrupole size which feels itself contracted. This implies an increase of frequency rotation.

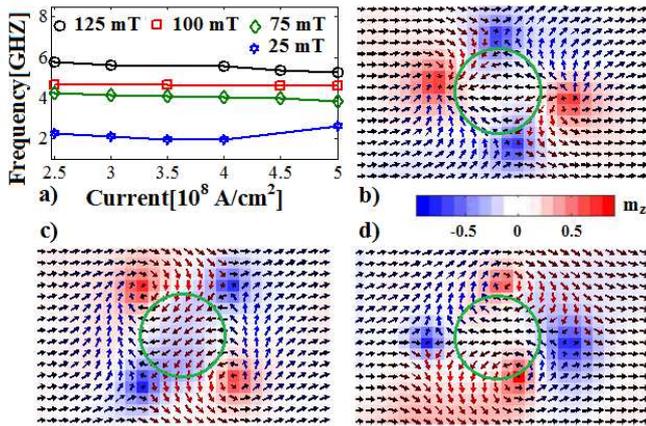

Fig. 4. The case of a rotating quadrupole in the rectangular section device of 500x300 nm$^2$. (a) Trend of frequency for values of applied field of 25 mT, 75 mT, 100 mT and 125 mT. (b,c,d) Configuration of the magnetization in different time instants (a zoomed-in area of the device is shown). Circular solid lines represent the contact area.

## IV. CONCLUSIONS

We have studied the vortex based soliton dynamics in different kinds of point-contact SVs. Depending on the shape and the size of the SV, we have achieved different configurations where the soliton mode is characterized by a rotating or translating vortex-antivortex pair, or a stable rotating quadripole composed by two vortex-antivortex pairs. The kind of dynamics depends also on the relative polarity of the vortex and the antivortex: when the polarities are opposite the pairs rotates, whereas, in the case of parallel polarities, they translate. Those results open new perspective in the dynamical soliton studies.

REFERENCES

[1] K. Y. Guslienko, B. A. Ivanov, V. Novosad, Y. Otani, H. Shima, and K. Fukamichi, "Eigenfrequencies of vortex state excitations in magnetic submicron-size disks," *J. Appl. Phys.*, vol. 91, pp. 8037-8039, May 2002.
[2] V. S. Pribiag, I. N. Krivorotov, G. D. Fuchs, P. M. Braganca, O. Ozatay, J. C. Sankey, D. C. Ralph, and R. A. Buhrman, "Magnetic vortex oscillator driven by d.c. spin-polarized current," *Nat. Phys.*, vol. 3, pp. 498-503, May 2007.
[3] A. Dussaux, B. Georges, J. Grollier, V. Cros, A.V. Khvalkovskiy, A. Fukushima, M. Konoto, H. Kubota, K. Yakushiji, and S. Yuasa, "Large microwave generation from current-driven magnetic vortex oscillators in magnetic tunnel junctions," *Nat. Comm.*, vol. 1, 8, April 2010.
[4] K. Yamada, S. Kasai, Y. Nakatani, K. Kobayashi, and T. Ono, "Current-induced switching of magnetic vortex core in ferromagnetic elliptical disks," *Appl. Phys. Lett.*, vol. 96, 192508, May 2010.
[5] H. Zhang, Y. Liu, M. Yan, R. Hertel, "Azimuthal Spin Wave Modes Excited in an Elliptical Nanomagnet With Vortex Pair States," *IEEE Trans Magn.*, vol. 46, pp. 1675-1678, June 2010.
[6] L.J. Chang, L. Pang, L.F. Shang, "Vortex induced by dc current in a circular magnetic spin valve nanopillar," *IEEE Trans. Magn.*, vol. 48, pp. 1297-1300, April 2012.
[7] G. Finocchio, V. Puliafito, S. Komineas, L. Torres, O. Ozatay, T. Hauet, and B. Azzerboni, "Nanoscale spintronic oscillators based on the excitation of confined soliton modes," *J. Appl. Phys.*, vol. 114, 163908, October 2013.
[8] V. Puliafito, L. Torres, O. Ozatay, T. Hauet, B. Azzerboni, and G. Finocchio, "Micromagnetic analysis of dynamical bubble-like solitons based on the time domain evolution of the topological density," *J. Appl. Phys.*, to be published.
[9] L. Berger, "Emission of spin-waves by a magnetic multilayer traversed by a current," *Phys. Rev. B*, vol. 54, pp. 9353-9358, October 1996.
[10] S.A. Wolf, "Spintronics: a spin-based electronics vision for the future," *Science*, vol. 294, pp. 1488-1495, November 2001.
[11] N. Locatelli, V. Cros, and J. Grollier, "Spin-torque building blocks," *Nat. Mat.*, vol. 13, pp. 11-20, January 2014.
[12] S. Parkin, X. Jiang, C. Kaiser, A. Panchula, K. Roche, and M. Samant, "Magnetically engineered spintronic sensors and memory," Proc. IEEE, vol. 91, pp. 661-680, May 2003.
[13] S. Bonetti, V. Puliafito, G. Consolo, V. Tiberkevich, A. Slavin, and J. Akerman, "Power and linewidth of propagating and localized modes in nanocontact spin-torque oscillators," *Phys. Rev. B*, vol. 85, 174427, May 2012.
[14] V. Puliafito, G. Consolo, L. Lopez-Diaz, and B. Azzerboni, "Synchronization of propagating spin-wave modes in a double-contact spin-torque oscillator: a micromagnetic study," *Phys. B*, vol. 435, pp. 44-49, February 2014.
[15] I.N. Krivorotov, D.V. Berkov, N.L. Gorn, N.C. Emley, J.C. Sankey, D.C. Ralph, and R.A. Buhrman, "Large-amplitude coherent spin waves excited by spin-polarized current in nanoscale spin valves," *Phys. Rev. B*, vol. 76, 024418, July 2007.
[16] D.V. Berkov, and N. L. Gorn, "Spin-torque driven magnetization dynamics in nanocontact setup for low external fields: Numerical simulation study," *Phys. Rev. B*, vol. 80, 064409, August 2009.
[17] L. Lopez-Diaz, D. Aurelio, L. Torres, E. Martinez, M.A. Hernandez-Lopez, J. Gomez, O. Alejos, M. Carpentieri, G. Finocchio, and G. Consolo, "Micromagnetic simulations using Graphics Processing Units," *J. Phys. D: Appl. Phys.*, vol. 45, 323001, July 2012.
[18] A. Giordano, G. Finocchio, L. Torres, M. Carpentieri, and B. Azzerboni, "Semi-implicit integration scheme for Landau-Lifshitz-Gilbert-Slonczewski equation," *J. Appl. Phys.*, vol. 111, 07D112, February 2012.
[19] G. Finocchio, O. Ozatay, L. Torres, R. A. Buhrman, D. C. Ralph, and B. Azzerboni, "Spin-torque-induced rotational dynamics of a magnetic vortex dipole," *Phys. Rev. B*, vol. 78, 174408, November 2008.
[20] S. Komineas, "Rotating vortex dipoles in ferromagnets," *Phys. Rev. Lett.*, vol. 99, 117202, September 2007.
[21] S. Komineas, "Frequency generation by a magnetic vortex-antivortex dipole in spin-polarized current," *Europhys. Lett.*, vol. 98, 57002, June 2012.
[22] V. L. Pokrovskii, and G. V. Uimin, "Dynamics of vortex pairs in a two dimensional magnetic material," *JETP Lett.*, vol. 41, pp. 105-108, February 1985.